\documentclass[aps,amssymb,showpacs,twocolumn,nofootinbib,floatfix]{revtex4}
\usepackage{graphicx}
\usepackage{epsfig}

\begin{document}
\title{Baryonic Pinching of Galactic Dark Matter Halos}

\author{Michael Gustafsson}
\email{michael@physto.se}
\author{Malcolm Fairbairn}
\email{malc@physto.se} \affiliation{Cosmology, Particle Astrophysics
and String Theory, Department of Physics, Stockholm University,
AlbaNova University Center, SE-106 91, Stockholm, Sweden}
\author{Jesper Sommer-Larsen}
\email{jslarsen@tac.dk} \affiliation{Dark Cosmology Centre, Niels
Bohr Institute, University of Copenhagen, Juliane Maries Vej 30,
DK-2100 Copenhagen, Denmark}

\date{December 21, 2006}

\begin{abstract}
High resolution cosmological N-body simulations of four galaxy-scale
dark matter halos are compared to corresponding
N-body/hydrodynamical simulations containing dark matter, stars and
gas. The simulations without baryons share features with others
described in the literature in that the dark matter density slope
continuously decreases towards the center, with a density
$\rho_{\rm{DM}} \propto r^{-1.3\pm0.2}$, at about 1\% of the virial
radius for our Milky Way sized galaxies. The central cusps in the
simulations which also contain baryons steepen significantly, to
$\rho_{\rm{DM}} \propto r^{-1.9\pm0.2}$, with an indication of the
inner logarithmic slope converging.
Models of adiabatic contraction of dark matter halos due to the
central build-up of stellar/gaseous galaxies are examined. The
simplest and most commonly used model, by Blumenthal \emph{et al.},
is shown to overestimate the central dark matter density
considerably. A modified model proposed by Gnedin \emph{et al.} is
tested and it is shown that while it is a considerable improvement
it is not perfect. Moreover it is found that the contraction
parameters in their model not only depend on the orbital structure
of the dark-matter--only halos but also on the stellar feedback
prescription which is most relevant for the baryonic distribution.
Implications for dark matter annihilation at the galactic center are
discussed and it is found that although our simulations show a
considerable reduced halo contraction as compared to the Blumenthal
\emph{et al.} model, the fluxes from dark matter annihilation is
still expected to be enhanced by at least a factor of a hundred as
compared to dark-matter--only halos.
Finally, it is shown that while dark-matter--only halos are
typically prolate, the dark matter halos containing baryons are
mildly oblate with minor-to-major axis ratios of $c/a=0.73\pm0.11$,
with their flattening aligned with the central baryonic disks.

\end{abstract}
\maketitle

\section{Introduction}\label{sec:intro}
There are still a multitude of open questions regarding the
distribution of dark matter in galactic halos, the effect of baryons
upon the structure of dark matter halos being one.  Many simulations
of galactic halos only take into account the dark matter (e.g.\
\cite{astro-ph-9508025,astro-ph-0311231,astro-ph-9903164,astro-ph-9908159,astro-ph-0306203,astro-ph-0312544})
, but one would expect the baryonic component of the galaxy to
behave very differently since it is able to cool (dissipate energy),
and contract considerably.  This is indeed observed in simulations
containing baryons and also in nature, where the baryons form a disk
and/or bulge at the center of apparently much more extended dark
matter halos.

It has long been realized that this ability of baryons to sink to
the center of galaxies would create an enhanced gravitational
potential well within which dark matter will congregate, increasing
the dark matter density there. To model this effect it is common to
use adiabatic invariants or some small modification of them
\cite{Eggen:1962dj,young,zeldovich,ASJOA.301.27,rydengunn,astro-ph-0406247,astro-ph-0501567,astro-ph-0507589,astro-ph-0601669,astro-ph-0604553}.
Such models are frequently used and it is of particular interest to
test the validity of these models
\cite{Barnes&White_84,astro-ph-0204164,astro-ph-0406247} currently
because of recent advances in mapping the velocity field of the
Milky Way \cite{astro-ph-0605025} and other galaxies. Furthermore,
upcoming gamma ray experiments will look for the flux due to the
self-annihilation of weakly interacting dark matter candidates from,
e.g., the galactic center.  Since this flux is proportional to the
dark matter density squared, predictions for, and conclusions from,
the data will depend strongly on the details of the effect of
baryonic pinching upon dark matter halos.

In this work we aim to investigate the effects of baryons on dark
matter halos in disk galaxies and test the most common models of
adiabatic contraction by comparing recent cosmological
N-body/hydrodynamical simulations of galaxies containing dark matter
and baryons to results from N-body dark-matter--only simulations of
the {\it same} halos. Fully cosmological simulations starting at
high redshifts are used. The simulation results are known from
previous studies to produce overall realistic gas and star
structures for spiral galaxies
\cite{astro-ph-0204366,astro-ph-0602595,astro-ph-0606531}, even
though the numerical resolution is still far from being able to
resolve any small scale features observed in real galaxies. The most
important dynamical property in this paper is the creation of stable
disk and bulge structures both for the gas and star components. The
\emph{angular momentum problem} is overcome by stellar feedback,
implying that the matter is not too centrally concentrated; a
generic problem of early galaxy simulations (see, e.g.,
\cite{astro-ph-0602351} and references therein on forming disk
galaxies in simulations). With the baryonic disks and bulges formed
fully dynamically the surrounding dark matter halo response should
also be realistically predicted inside the simulated spiral
galaxies.

\smallskip
The paper is organized as follows. In section \ref{sec:sim} we
present the simulations, and then focus in section
\ref{sec:DM_profile}, in particular, upon profile fits to the
density of the dark matter halos in the simulations with and without
baryons. In section \ref{sec:adiabatic} we investigate different
prescriptions which aim to predict the effects of baryons upon dark
matter profiles by testing whether they are able to reproduce the
dark matter profiles observed in the simulations with baryons. We
then comment in section \ref{sec:WIMP} on how these results might
change the expected flux of gamma rays from dark matter annihilation
in spiral galaxies, including the Milky Way. Finally, we analyze in
section \ref{sec:non-sph} the nonsphericity of the different
components of the simulations -- dark matter, gas and stars -- to
attempt to further quantify their effects upon each other before we
summarize our results in section \ref{sec:summary}.

\section{Simulations}\label{sec:sim}

\begin{table*}
\begin{center}
 \caption[]{\it Main properties of dark matter halos and galaxies at $z$=0}\vspace{0.0cm}
\begin{tabular*}{\textwidth}{@{\extracolsep{\fill}}lllllllll}
\hline\hline
 Simulation                                               &  S1  & DM1 & S2  & DM2 & S3  & DM3 & S4  & DM4  \vspace*{0.05cm}\\ \hline
Virial radius $r_{200}$ [kpc]                             & 209  & 211 & 200 & 201 & 100 & 102 & 98.3& 97.5 \\
Total mass $M_{200}$ [$10^{11} M_\odot$]                  & 8.9  & 9.3 & 7.8 & 8.0 & 1.0 & 1.1 & 0.93& 0.91 \\
Number of particles $N_{200}$ [$\times 10^5$]             & 3.6  & 1.2 & 3.5 & 1.0 & 3.2 & 0.98& 3.1 & 0.90 \\
DM particle mass $m_{\rm{dm}}$ [$10^6 M_\odot$]                & 6.5  & 7.6 & 6.5 & 7.6 & 0.81& 0.95& 0.81& 0.95 \\
SPH particle mass $m_{\rm{baryon}}$ [$10^6 M_\odot$]           & 1.1  &\ldots & 1.1 &\ldots & 0.14&\ldots & 0.14&\ldots  \\
Grav. soft length DM $\epsilon_{\rm{dm}}$ [kpc]                & 1.0  & 1.0 & 1.0 & 1.0 & 0.5 & 0.5 & 0.5 & 0.5  \\
Grav. soft length SPH $\epsilon_{\rm{baryon}}$ [kpc]           & 0.6  &\ldots & 0.6 &\ldots & 0.3 &\ldots & 0.3 &\ldots  \\
Characteristic circular  speed $V_{\rm{c}} [km/s]$
\footnote{$V_{\rm{c}}$ is determined as in \cite{astro-ph-0204366}}
                                                          & 245  &\ldots& 233 &\ldots & 124 &\ldots & 122 &\ldots\\
Specific angular momentum (bulge+disk), $j_*$ [kpc km/s]  & 447  &\ldots& 303 &\ldots & 144 &\ldots & 153 &\ldots\\
Specific angular momentum (cold gas), $j_{\rm{cg}}$ [kpc km/s] & 1895 &\ldots & 1055&\ldots &1005 &\ldots & 1093&\ldots\\
Star formation rate (SFR)  [$M_\odot/yr$]                 & 1.4  &\ldots& 1.7 &\ldots & 0.13&\ldots & 0.15&\ldots\\
 $b = SFR/\langle SFR \rangle$                            & 0.23 &\ldots& 0.32&\ldots & 0.13&\ldots & 0.16&\ldots\\
Baryonic disk + bulge mass [$10^{10} M_\odot$]            & 7.17 &\ldots& 5.79&\ldots & 1.13&\ldots & 0.98&\ldots  \\
 Baryonic Bulge-to-disk mass ratio                        &0.19  &\ldots& 0.80&\ldots & 0.76&\ldots & 0.60&\ldots \\\hline\hline
\end{tabular*}
\label{tab:sim}
\end{center}
\end{table*}

The simulated galaxies used in our work consist of two Milky Way
sized galaxies with virial radii (at $z$=0) of $r_{200}\approx$ 200
kpc and two smaller galaxies with virial radii of around 100 kpc.
The two larger galaxies are labeled 1 and 2 and the two smaller
galaxies 3 and 4; the number simply being an identifying label. Here
we have followed common practice and defined $r_{200}$ as the radius
of the sphere enclosing the mass $M_{200}$ within which the mean
density is 200 times the critical density, $\rho_c=3H_0^2/8\pi G$.

All four galaxies have been extracted from fully cosmological
simulations using the \textsc{Hydra} code and an improved version of
the Smoothed Particle Hydrodynamics code \textsc{TreeSPH}. Since the
software generating the disk galaxies has been used in many previous
works (see,
e.g.,\cite{astro-ph-0204366,astro-ph-0602595,astro-ph-0606531} ), we
will only briefly mention the main features of the numerical code.

The simulations are performed in a $\Lambda CDM$ cosmology with
$\Omega_M=0.3$, $\Omega_\Lambda=0.7$, $H_0=100h$km s$^{-1}$Mpc$^{-1}
=65$km s$^{-1}$Mpc$^{-1}$ and with the matter power spectra
normalized such that the present linear root mean square (rms)
amplitude of mass fluctuations inside 8$h^{-1}$Mpc is
$\sigma_8=1.0$. There is still some uncertainty in the accepted
values of $\sigma_8$ and $h$, but we have tested that none of the
general conclusions obtained in this paper are affected by small
changes in these parameters.

The galaxies are generated by first performing a dark-matter--only
simulation, using the \textsc{Hydra} code, with 128$^3$ particles in
a box of comoving length of 10 $h^{-1}$ Mpc and starting at redshift
$z_i=39$. After running this simulation, galactic size objects are
identified.  The simulations which include baryons (or alternatively
with only dark matter) are then set up, particles within
4\,$r_{\rm{vir}}$ at $z=0$ are traced back to their initial
conditions, the dark matter particle mass resolution is increased by
up to a factor of 64, and one SPH particle, i.e.\ baryonic matter,
per dark matter particle is added (keeping the total mass and fixing
the baryonic fraction $f$ to 0.15). The simulations are then rerun
with the improved \textsc{TreeSPH} code; incorporating star
formation, stellar feedback processes, radiative cooling and
heating, etc. The final result are qualitatively similar to observed
disk and elliptical galaxies at $z=0$, a result which is mainly
possible by overcoming the \emph{angular momentum problem} by an
early epoch of strong, stellar energy feedback in form of SNII
energy being fed back to the intrastellar medium.

To study the effect of baryons on dark matter halos in Milky Way
like galaxies we use four simulated disk galaxies with the highest
available resolution at our disposal.  By comparing simulations with
different resolutions we infer that the results are robust down to
an inner radius $r_{\rm{min}}$, which is 2 times the gravitational
softening length. This is in approximate agreement also with other
commonly used convergence criteria, such as $r_{\rm{min}} \approx
N_{200}^{-1/3} r_{200}$ (see, e.g.,
\cite{Power:2002sw,astro-ph-0312544} and references therein). For
the larger galaxies 1 \& 2 we deduce $r_{\rm{min}} = 2$kpc whereas
for the two smaller 3 \& 4 $r_{\rm{min}} = 1$kpc.  The spiral
galaxies containing dark matter and baryons will be labeled with 'S'
(e.g.\ S1) whereas those containing only dark matter will be labeled
with 'DM' (e.g.\ DM3). There are some notable differences between
the four galaxies, for example the large galaxy S1 has a very
pronounced flat gas and stellar disk with a star bulge, whereas
galaxy S2 is strongly barred. The gas in the two smaller galaxies
has a very definite disk structure and the stars exhibit both disks
and central bulges which are more centrally concentrated than the
ones in the two larger galaxies.

A summary of the main parameters of the simulated galaxies is given
in Table~\ref{tab:sim} (see
\cite{astro-ph-0204366,astro-ph-0602595,astro-ph-0606531} for
further details).

\section{Dark matter halo radial profiles}\label{sec:DM_profile}

\begin{figure*}
\centerline{\epsfig{file=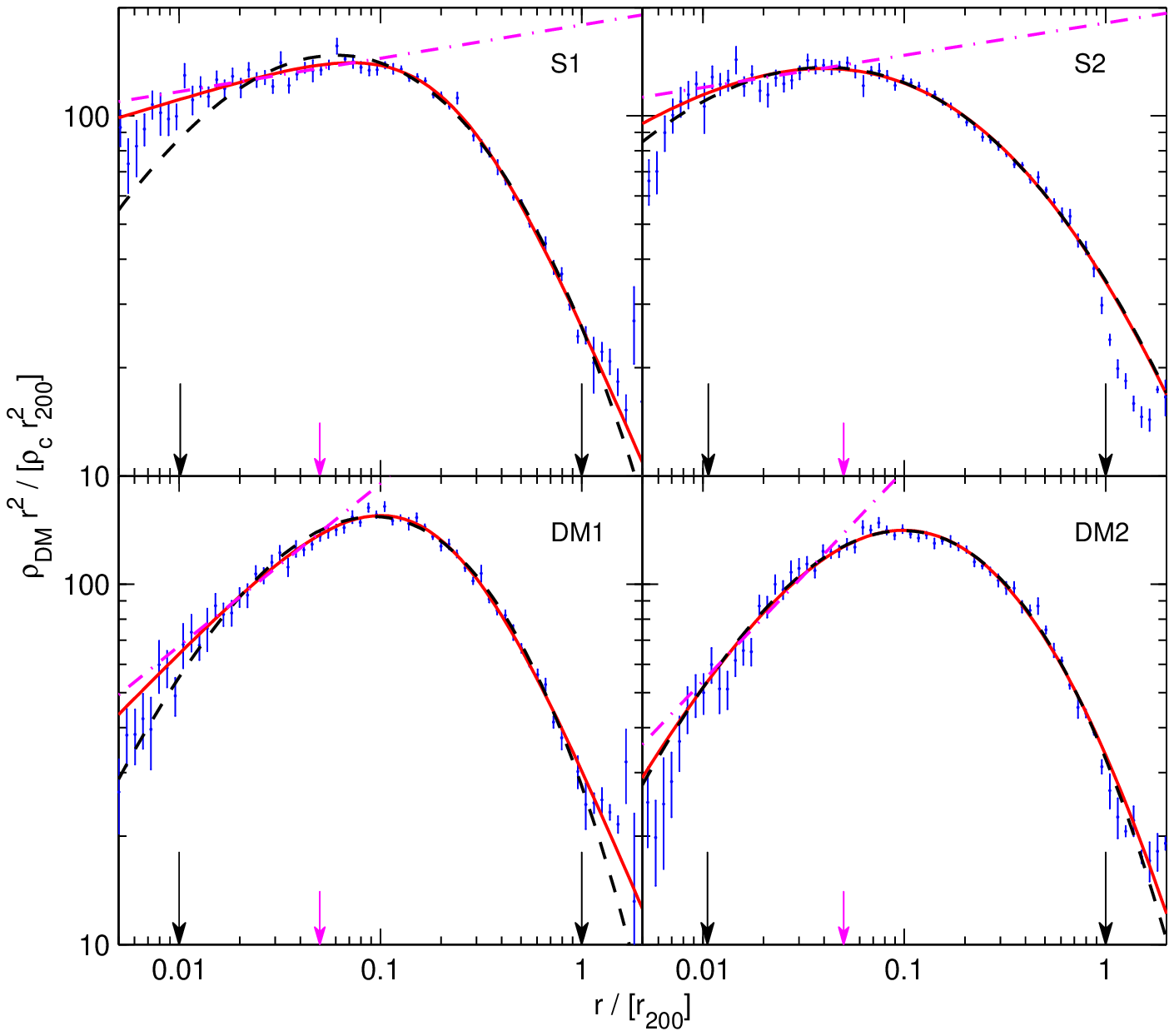,width=1.8\columnwidth}}
\caption{\it
 Top panels: dark matter density profiles from galaxy
simulations including baryons; galaxy S1 (left) and S2 (right).
 Bottom panels: density profiles from halo simulations including
only dark matter;  halo DM1 (left) and DM2 (right).
 The solid, dashed and dot-dashed curves show the best fit from the parametrization given in Eq.~(\ref{eq:abg}), (\ref{eq:NFW_new}) and a single power law, respectively.
 Long arrows show the lower resolution limit ($r_{\rm{min}}$) and the
virial radius ($r_{200}$), respectively. The shorter arrows indicate
the upper limit for the single power law fits (0.05 $r_{200}$) .}
\label{density12}
\end{figure*}

\begin{figure*}
\centerline{\epsfig{file=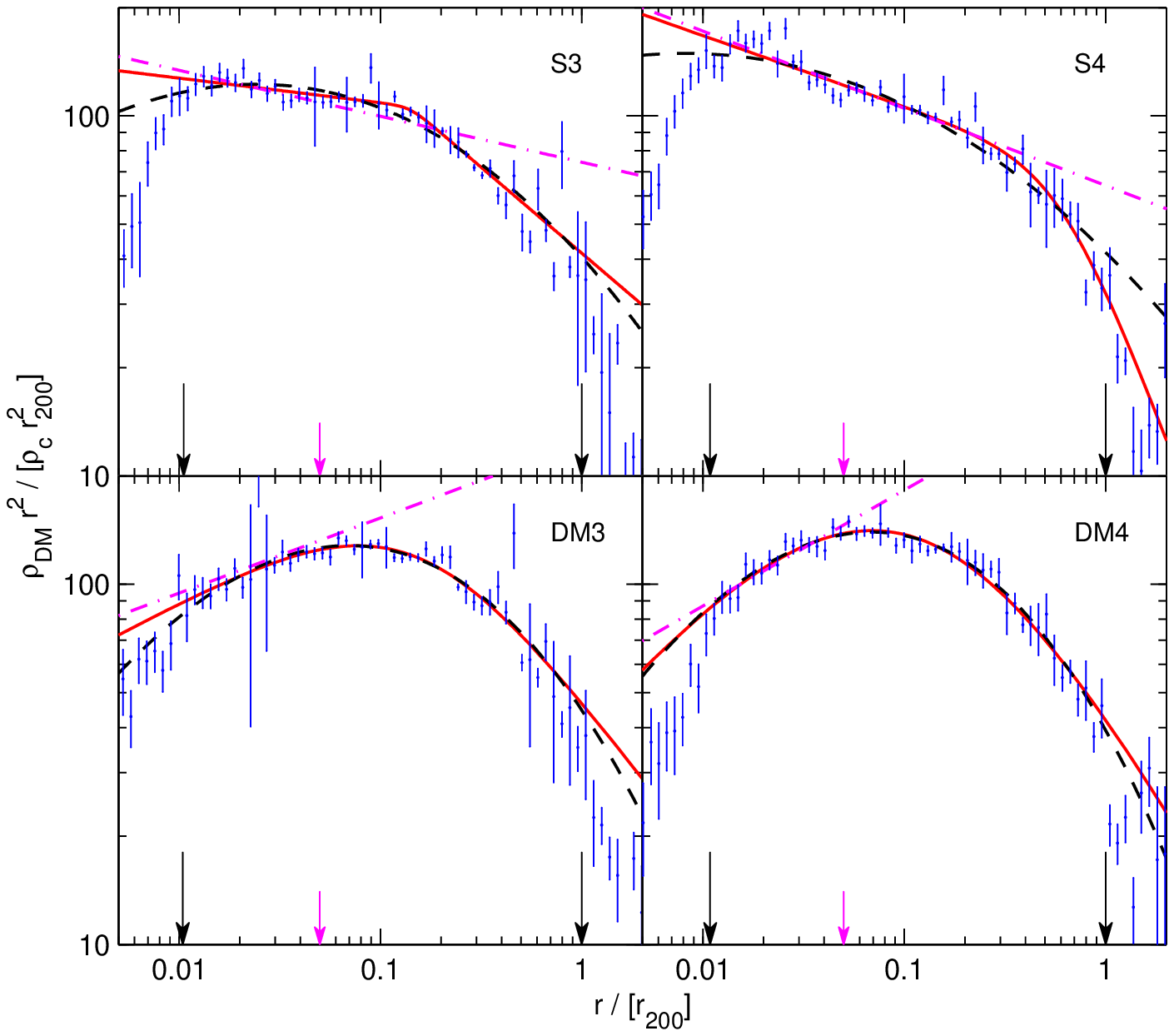,width=1.8\columnwidth}}
\caption{\it
 Top panels: dark matter density profiles from galaxy
simulations including baryons; halo S3 (left) and S4 (right).
 Bottom panels: density profiles from halo simulations including
only dark matter;  halo DM3 (left) and DM4 (right).
 The solid, dashed and dot-dashed curves show the best fit from the parametrization given in Eq.~(\ref{eq:abg}), (\ref{eq:NFW_new}) and a single power law, respectively.
 Long arrows shows the lower resolution limit ($r_{\rm{min}}$) and the virial radius
($r_{200}$), respectively. The shorter arrows indicate the upper
limit for the single power law fits (0.05 $r_{200}$) .}
\label{density34}
\end{figure*}

In the section on nonsphericity we will address the nonspherical
aspects of the simulations in more detail, in particular the effect
of the triaxiality of the dark matter halo upon the baryons and
vice-versa. However before that, in order to directly compare the
profiles with most others in the literature, we assume spherical
symmetry and fit the galaxies with common radial profiles.  One
parametrization which assumes two asymptotic radial power law
behaviors at both small ($\gamma$) and large ($\beta$) radii is
known as the '$\alpha\beta\gamma$' profile (or the Zhao profile),
where the density as a function of radius is given by the expression
\begin{equation}
\rho(r)=\frac{\rho_0}{(r/r_s)^\gamma\left[1+(r/r_s)^\alpha\right]^{\frac{\beta-\gamma}{\alpha}}}
\label{eq:abg}
\end{equation}
where $\alpha$ governs the radial rate at which the profile interpolates
between the asymptotic powers $-\gamma$ and $-\beta$.

Often in the literature various constraints on these parameters are
assumed, resulting in subclasses of less general profiles.  In
particular the density profile in the outer part of the galaxy halo,
$\beta$, is often assumed to be 3, which is the canonical value
found for halos in simulations invoking only dark matter.  For this
reason many authors use $(\alpha,\beta,\gamma)=(1,3,\gamma)$ and is
commonly known as the generalized Navarro, Frenk and White (NFW)
profile (with the standard NFW profile having $\gamma=1$)
\cite{astro-ph-9508025}. We in general find a significantly better
fit to our profiles if we also leave $\alpha$ and $\beta$ as free
parameters. Moreover, it is obviously not certain that profiles used
to characterize the halos of dark-matter--only simulations should
also give good fits to the dark matter halos formed in simulations
also containing baryons.
We have therefore done least $\chi^2$ fits to the profiles leaving
the four parameters in Eq.~(\ref{eq:abg}) free ($\rho_0$ we always
constrain by the total mass in the fitting range which we set to be
between $r_{\rm{min}}$ and $r_{200}$). The profile fits can be seen
in Fig.~\ref{density12}, \ref{density34} as solid lines and the
parameter values are given in Table~\ref{tab:abg}.

It is useful to keep in mind that with four free parameters in the
'$\alpha\beta\gamma$' profile, there are some degeneracies in the
inferred parameter values \cite{klypin2001}. The numbers given in
Table~\ref{tab:abg} aim to give a good parametrization of the
density profiles in the fitted range, and do not neccesarily claim
to represent profiles that could be extrapolated into smaller radii
with confidence.

\begin{table*}[t]
\begin{center}
 \caption{\it Best fit parameters for model Eq.~(\ref{eq:abg}) to the spherical symmetrized dark matter halos.}\vspace{0.0cm}
\begin{tabular*}{0.9\textwidth}{@{\extracolsep{\fill}} l@{\hspace{20pt}} l@{\hspace{10pt}} l@{\hspace{10pt}} l@{\hspace{10pt}} l@{\hspace{10pt}} l}\hline\hline
 Galaxy sim. with(without) baryons            & $r_{s} [kpc]$ & $\alpha$ & $\beta$ & $\gamma$ & $\chi^2_{dof}$ [46 dof]\\
\hline
 S1 (DM1)  & 44.9 (36.0)& 1.76  (1.47)   & 3.31 (3.36) & 1.83 (1.42)  & 1.4 (1.0) \\
 S2 (DM2)  & 150  (85.1)& 0.486 (0.588)  & 4.21 (4.79) & 1.49 (0.850) & 1.5 (1.2)\\
 S3 (DM3)  & 13.3 (14.3)& 12.4  (1.387)  & 2.48 (2.74) & 2.07 (1.70)  & 1.0 (2.1)\\
 S4 (DM4)  & 56.1 (10.3)& 2.75  (0.915)  & 3.48 (3.00) & 2.20 (1.36)  & 2.1 (1.0)\\
\hline\hline
\end{tabular*}
 \label{tab:abg}
\end{center}
\end{table*}

We have also fitted the dark matter halo density distributions over
the same radial range using exponential profiles, where the
logarithmic slope changes continuously with radius as recently
suggested by Navarro, Frenk and White \cite{astro-ph-0311231}
\begin{equation}
 \rho(r) = \rho_{-2}\,\exp\left[-\frac{2}{\alpha}\left(\left(\frac{r}{r_{-2}}\right)^\alpha-1\right)\right].
\label{eq:NFW_new}
\end{equation}
In this profile, $\rho_{-2}$ and $r_{-2}$ correspond to the density
and radius where $\rho\propto r^{-2}$. The best fit values can be
found in Table~\ref{tab:NFW_new} and the profile fits can be seen in
Fig.~\ref{density12}, \ref{density34} as dashed lines.

\begin{table*}[t]
 \caption{\it Best fit parameters for model Eq.~(\ref{eq:NFW_new}) to the spherical symmetrized dark matter halos.}\vspace{0.0cm}
\begin{tabular*}{0.75\textwidth}{@{\extracolsep{\fill}} llll}\hline\hline
 Galaxy sim. with(without) baryons            & $r_{-2} [kpc]$ & $\alpha$ & $\chi^2_{dof}$ [48 dof]\\ \hline
 S1 (DM1)  & 11.9  (18.5)  & 0.185  (0.247)  & 3.1 (1.5)  \\
 S2 (DM2)  & 7.66  (18.7)  & 0.117  (0.226)  & 1.5 (1.5)  \\
 S3 (DM3)  & 2.30  (6.70)  & 0.0728 (0.132)  & 2.6 (2.1)  \\
 S4 (DM4)  & 0.750 (6.31)  & 0.0507 (0.153)  & 3.2 (0.91) \\
\hline\hline
\end{tabular*}
 \label{tab:NFW_new}
\end{table*}

Finally, we have fitted single power laws to the central dark matter
density profile (r$_{\rm{min}}<r<0.05 r_{200}$) to determine the
averaged logarithmic slope of the resolved central cusp. These slope
values are found in Table~\ref{tab:innerslope}. This can also partly
be compared to the central asymptotic logarithmic slope in
Eq.~(\ref{eq:abg}), which is $-\gamma$ (whereas in the profile in
(\ref{eq:NFW_new}) the logarithmic slope is continuously decreasing
towards the center).

\smallskip
To describe the dark matter halos we actually fit
$\textrm{d}m/\textrm{d}r=4\pi \rho r^2$ since this is representative
of the actual data in the simulation.  Logarithmic binning of the
radius is used to minimize the effect of substructure, which becomes
more important at large radii, while at the same time capturing the
behavior of density profile in the central part.

In order to obtain an estimate of the variance with which to do the
fits and for our subsequent analyses throughout the paper, we take
five snapshots of each simulation at different times.  These
snapshots correspond to today's epoch and to four successive earlier
times with $\Delta t=200$ Myr, and from these we find their one
standard deviation (unbiased) mean square dispersion. This $\Delta
t$ is long enough so that the particles in the inner regions of the
simulations will have had time to completely change their positions
with respect to the center of the galaxy, but a short enough
timescale in total so that the overall mass profile of the halo has
changed very little (for all galaxies, the dark matter mass within
50 kpc increases by less than about 1\% over the total period of 1
Gyr).

This method of estimating the variance in each bin has the effect of
suppressing the influence from temporary small scale inhomogeneities
in density due to dark matter subhalos (which are most common at
large radii) and at the same time retaining the Poisson population
variance in other bins. It turns out that the variances are
approximately equal to the Poissonian values for most radial bins.
This will serve as the estimate of the uncertainties in our
subsequent investigations. A more detailed analysis of the actual
uncertainties which would take into account, e.g.\ systematic radial
dependencies from numerical and resolution effects or even effects
from the implementation of physical processes themselves, is very
difficult to achieve and is beyond the scope of this paper. From the
five time frames no strong correlations between our bins were found
and we will in practice not take into account such eventual
correlations.

Tables \ref{tab:abg}-\ref{tab:innerslope} contain significant
information:  For the '$\alpha\beta\gamma$' profiles one has
$\rho\propto r^{-\beta}$ at large radii, and even when $\beta$ is
left as a free parameter, values for $\beta$ around 3 emerge, albeit
with a rather large scatter. Moreover, the $\chi^2$ per degree of
freedom is significantly smaller when using the
'$\alpha\beta\gamma$' profiles rather than the exponential profiles
for the simulation with baryons.

From the best fit parameters in Table~\ref{tab:abg} it follows that
the asymptotic central logarithmic slope (-$\gamma$) for the
dark-matter--only simulations average to -1.1 for the two larger
halos DM1 and DM2 and a somewhat steeper slope of -1.5 for the two
smaller galaxies. These results are, to the inner resolved radii, in
agreement with other recent simulations including only dark matter
(see e.g.\ \cite{astro-ph-0504215} and references therein).

The effect that the presence of baryons has upon the central slope
of the dark matter density profile is very pronounced, pushing the
average asymptotical central logarithmic slope up to -1.7 for the
larger galaxies and to -2.1 for the two smaller galaxies. Another
way of seeing this systematic steepening of the profiles due to the
presence of baryons is to look at the fits to the exponential
profile (\ref{eq:NFW_new}) presented in Table~\ref{tab:NFW_new}. For
all four galaxies, the presence of baryons brings in the radius at
which the density is dropping as $\rho\propto r^{-2}$ (for the
smaller galaxies, this radius becomes comparable to $r_{\rm{min}}$).

\begin{figure}
\centerline{\epsfig{file=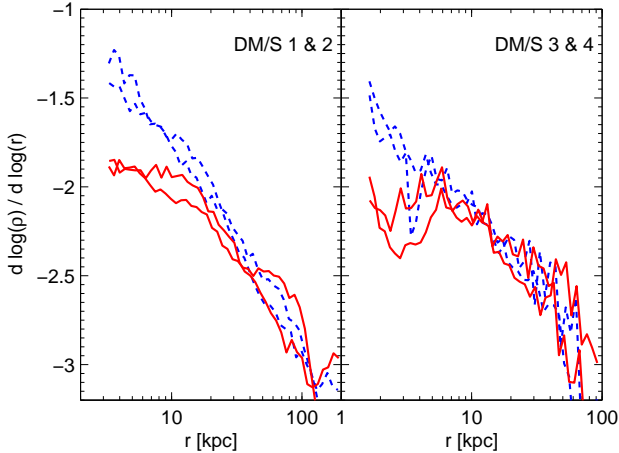,width=1\columnwidth}}
\caption{\it Logarithmic slope of the dark matter density profiles
for all halos, plotted versus radius. Solid/red curves from the
simulation including baryons, and dashed/blue curves from the
simulations including only dark matter.
 Left panel: the larger, Milky Way sized, galaxies (1 \& 2).
 Right panel: the two smaller galaxies (3 \& 4).} \label{fig:slope}
\end{figure}

\begin{table}
 \caption[]{\it Best fit parameters for single power law $\rho\propto r^{-\gamma}$ fits to the central dark matter density profiles ($r_{\rm{min}}<r<0.05r_{200}$).}\vspace{0.0cm}
\begin{tabular*}{\columnwidth}{@{\extracolsep{\fill}} lll}\hline\hline
 Galaxy sim.             & $\gamma$ & $\chi^2_{dof}$ \\
 with(without) baryons  &          & [49 dof]       \\\hline
 S1 (DM1)  & 1.91 (1.56)  & 0.65 (1.2)  \\
 S2 (DM2)  & 1.91 (1.41)  & 0.76 (1.1)  \\
 S3 (DM3)  & 2.13 (1.79)  & 0.90 (0.92) \\
 S4 (DM4)  & 2.21 (1.68)  & 1.6\,\,\,  (0.80) \\
\hline\hline
\end{tabular*}
\label{tab:innerslope}
\end{table}

Since all simulations have limited resolution one should be cautious
in extrapolating profiles inside the resolved radius of the
simulations. To explore the inner slope further we show in
Fig.~\ref{fig:slope} the logarithmic slope $\textrm{d} \log \rho
/\textrm{d}\log r$. Note from this figure that there is no sign of
convergence of the central slope in the simulations without baryons,
whereas in the simulation including baryons the slope change is
drastically less and might already have converged to a defined
value. To calculate the logarithmic slope profiles without excessive
particle noise we average both over the five time frames and over
the five nearest radial neighbor bins (corresponding to logarithmic
smearing window of 20\% of our logarithmic radius range
$r_{\rm{min}}<r<r_{200}$). This eliminates most of the fluctuations
without biasing the slope significantly.

For the two larger galaxies Fig.~\ref{fig:slope} clearly
demonstrates that the logarithmic slope is always continuously
changing for the dark-matter--only simulations, while for the
simulations including baryons the logarithmic slope derivative is
drastically less in the central regions. This may indicate that the
inner logarithmic slope converges to a value close to -1.9, i.e.\
close to an isothermal sphere. The picture is slightly less clear
for the two smaller galaxies, but there the logarithmic slope also
flattens out in the inner region to a value roughly around -2.1
(although the bin-to-bin scatter is somewhat too large to draw any
firm conclusions).

If we only fit the central part of the dark matter density
($r_{\rm{min}} < r < 0.05 r_{200}$) a single power law fit should
work well for the simulation including baryons and we deduce from
Table~\ref{tab:innerslope} a inner logarithmic slope close to -1.9
for the two larger galaxies and an average inner slope of -2.2 for
the two smaller galaxies.

Galaxy S2 is different compared to the other galaxies in several
ways. For instance, for the dark-matter--only simulation of the halo
(DM2), the best fit to the '$\alpha\beta\gamma$' density profile
yields a asymptotic central logarithmic slope of only
$-\gamma=-0.85$ and a very large scale radius $r_s$. One thing that
separate this galaxy from the others in its evolution is that it has
experienced a late time merger and that there is not as strong disk
structure, but rather a pronounced bar structure in the stars.
Despite this the dark matter density profile is not entirely
different, which seems to support the results reported in
\cite{astro-ph-0510583} -- where they find that dark matter profile
shapes are preserved in mergers.

\begin{figure}
\vspace{4.2mm}
\centerline{\epsfig{file=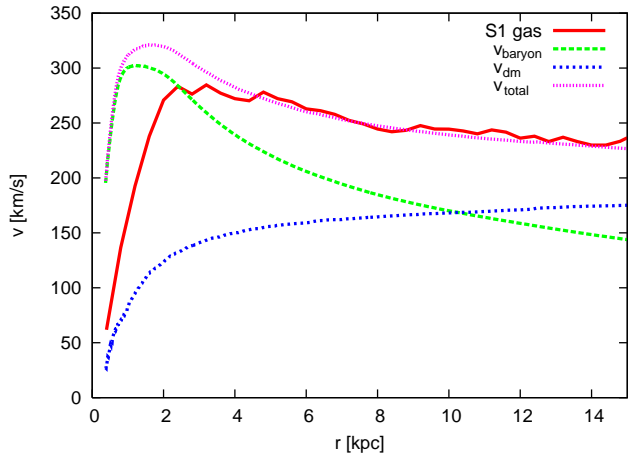,angle=270,width=0.95\columnwidth}}
\caption{\it Rotation curve of the disk gas for galaxy S1 (solid red
curve). Also plotted are the Keplerian velocities expected due to
baryons, dark matter and in total, assuming spherical symmetry
($v_{x}=\sqrt{GM_{x}(\!<\!r)/r}$).} \label{rot15}
\end{figure}

The most obvious comparison with observations, concerning dynamics
from dark matter, is the rotation curves. In Fig.~\ref{rot15} we
plot the rotation curve for the gas in one of the larger galaxies
(S1). The solid red line is the average in each radial bin of the
magnitude of vectors $\vec{r}\times \vec{v}/|\vec{r}|$ of which
there is one for each cold disk gas particle. The dashed lines
correspond to the Keplerian velocities expected at each radius due
to dark matter, baryons and the sum of the two.

Even though the dark matter distribution in the core of the
simulation seems to asymptote to a cusp as far as it is possible to
ascertain above the smallest length scale resolved, it turns out
that the baryonic mass in the center of the galaxy is predominantly
responsible for the rotation of the inner part of the gas disk.  The
rotation curve is comparable to observed rotation curves of large
disk galaxies
\cite{astro-ph-9506004,astro-ph-0010594,astro-ph-0408132}.

The rotation curves in the simulations of the two smaller galaxies
are somewhat too centrally peaked compared to most observations.
This is probably related to the fact that these simulated galaxies
don't fully overcome the \emph{angular momentum problem} in the
inner couple of kpc, and hence that the central baryonic component
of these galaxies might still be somewhat too concentrated.

In the inner couple of kpc there is a discrepancy between the actual
circular velocity of the gas and the total rotational velocity
expected from the enclosed mass. This is mainly a numerical effect,
due to gravity softening, but also partly due to effects of
noncircular motions of, and pressure gradients in, the cold gas, as
discussed by \cite{astro-ph-0509644}.

\section{Testing Adiabatic Contraction}\label{sec:adiabatic}

The most commonly used model of baryonic contraction was suggested
by Blumenthal \emph{et al.} \cite{ASJOA.301.27} and is based on two
assumptions, namely that the orbits of particles are circular and
that the dark matter halo contracts adiabatically, i.e.\ slowly,
compared to the dynamical time scale of the system. Consider a dark
matter halo which has an initial mass profile $M_i(r)$. One can then
ask what the effect would be of changing a fraction $f$ of those
particles into baryons. The dissipational baryons will cool and
contract, and end up with a final mass distribution $M_b(r)$.  In
the adiabatic contraction model of Blumenthal \emph{et al.}, the
relationship between the initial mass profile and the final dark
matter profile $M_x(r)$ is given by
\begin{equation}
r\left[M_b(r)+M_x(r)\right]=r_iM_i(r_i)=r_iM_x(r)/(1-f)
\label{eq:old}
\end{equation}
which relies upon the assumption of conservation of angular
momentum, a spherically symmetric gravitational potential, and the
noncrossing of the circular orbits during the contraction, a
criterion which gives $(1-f)M_i(r_i)=M_x(r)$.

Since this proposal was put forward, it has been established that
typical orbits in N-body simulations of dark matter halos are rather
elliptical (see, e.g., \cite{astro-ph-9801192}) so that $M(r)$
changes around the orbit and $M(r)r$ is no longer an adiabatic
invariant. It has therefore been pointed out by Gnedin \emph{et al.}
\cite{astro-ph-0406247} that the relation in Eq.~(\ref{eq:old})
could be modified to try to take this into account. In particular
they argue that using the value of the mass within the average
radius of a given orbit, $\bar{r}$, should give better results than
that within the instantaneous radius $r$ or the maximum radius at
apogee $r_a$. The average radius $\bar{r}$ for a particle is given
by
\begin{equation}
 \bar{r}=\frac{2}{T}\int_{r_p}^{r_a}\frac{r}{v_r}\textrm{d}r \label{eq:rbar}
\end{equation}
where $v_r$ is the radial velocity, $r_p$ is the perihelion radius
and $T$ is the radial period. The ratio between $r$ and $\bar{r}$
will change throughout the halo so Gnedin \emph{et al.} parameterize
the average $\langle\bar{r}\rangle$ ($\bar{r}$ averaged over the
population of orbits at a given $r$) using the power law with two
free parameters
\begin{equation}
\langle \bar{r}\rangle=r_{200}A\left(\frac{r}{r_{200}}\right)^w
\label{eq:poly}
\end{equation}
For their simulations, they find values of $A=0.85\pm0.05$,
$w=0.8\pm0.02$ (Note that their definition of the virial radius is
$r_{180}$ rather than $r_{200}$ as used in this paper, but the
effect of this difference upon $A$ is very small;
$A_{180}=A_{200}(r_{200}/r_{180})^{1-w}$.)

\begin{figure}
\vspace{4.2mm}
\centerline{\epsfig{file=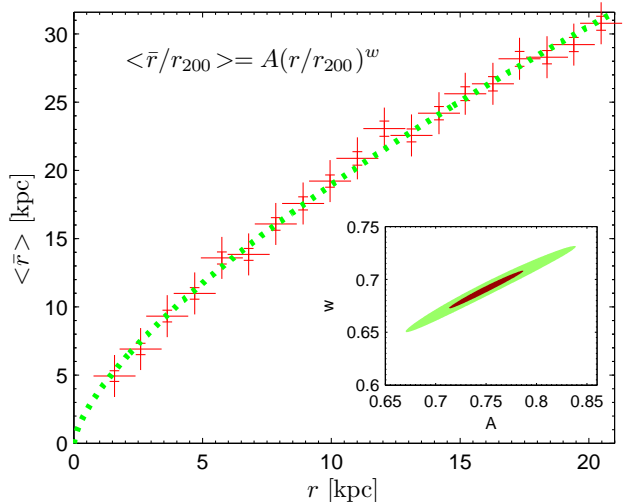,width=0.95\columnwidth}}
\caption{\it Points are $\langle\bar{r}\rangle$ vs. $r$ for the halo
DM1.  The best fit relation, corresponding to ($A$,
$w$)=(0.74,0.69), shows that the power law assumption
(\ref{eq:poly}) is an excellent representation of the data.  The red
crosses represent the binned data and the smaller horizontal red
lines represent the vertical variance on that data. The smaller
figure shows in red (darker shading) the 1$\sigma$ (68\%) confidence region
whereas the green (lighter shading) area is the 3$\sigma$ (99.7\%) confidence
region in the ($A$,$w$) plane.} \label{power}
\end{figure}

This power law assumption for the relationship between $r$ and
$\langle\bar{r}\rangle$ in Eq.~(\ref{eq:poly}) turns out to be a
good model of the orbital structure of all our halos, as illustrated
by, e.g., the DM1 halo in Fig.~\ref{power} , where
$\langle\bar{r}\rangle$ is plotted as a function of $r$ with a power
law fit running through the data. We do not actually perform the
integral in Eq.~(\ref{eq:rbar}), but rather take the average of the
radii over the 5 time snapshots and bin radially, and fit to the
average within each bin, to determine $A$ and $w$. To calculate the
($A$,$w$) parameters a $\chi^2$ fit is performed in 19 radial bins
between $r_{\rm{min}}$ and 0.1${r_{200}}$ and using the standard
deviation of the average of $\bar{r}$ in each radial bin. The
results are presented in Table~\ref{tab:Aw}.

\begin{table}
\caption[]{\it Values of $A$ and $w$ in Eq.~(\ref{eq:rbar}) by
fitting $\langle\bar{r}\rangle$ as a function of $r$.  The ranges
stated are the joint 1$\sigma$ intervals.}\vspace{0.0cm}
\begin{tabular*}{\columnwidth}{@{\extracolsep{\fill}} l@{\hspace{20pt}} l@{\hspace{7pt}} l@{\hspace{7pt}} l@{\hspace{20pt}} l@{\hspace{7pt}} l@{\hspace{7pt}} l}\hline\hline
 DM halo& $A_{\rm{min}}$ & $A_{best}$& $A_{max}$ & $w_{\rm{min}}$& $w_{best}$&$w_{max}$ \\
\hline
 DM1  & 0.71 & 0.74 & 0.79 & 0.67 & 0.69 & 0.71\\
 DM2  & 0.78 & 0.83 & 0.88 & 0.69 & 0.71 & 0.73\\
 DM3  & 0.72 & 0.76 & 0.81 & 0.80 & 0.81 & 0.83\\
 DM4  & 0.69 & 0.74 & 0.78 & 0.79 & 0.80 & 0.82\\
\hline\hline
\end{tabular*}
\label{tab:Aw}
\end{table}

In order to test the hypothesis of Gnedin \emph{et al.} we first
determine the best fit values of $A$ and $w$ directly from the
orbital structure (or in other words the dark matter
``ellipticity''), as discussed above. We then perform baryonic
contraction of the dark matter in the simulations {\it without}
baryons by scanning over different $A$ and $w$ (see below), to test
whether the values required to reconstruct the simulations
\emph{with baryons} correspond to those found directly from the
orbital ellipticities.

\smallskip
For each galaxy, we label the halo resulting in the
dark-matter--only simulation by '$DM$' so that the mass inside
radius $r$ of that halo is denoted by $M^{\rm{DM}}(r)$. The second
simulation, including dark matter and baryons, we denote by '$S$'.
We thus label the mass profile of the baryons by $M_{b}^S(r)$,
whereas the mass profile for the dark matter that will be predicted
by the Gnedin \emph{et al.} model \cite{astro-ph-0406247} will be
denoted by $M_{\rm{dm}}^X(r)$. The predicted density profile
$M_{\rm{dm}}^X(r)$ is then compared to what is actually found from
our simulation including baryons, which we denote
$M_{\rm{dm}}^S(r)$. To run the simulations with and without baryons
with the same total mass it is necessary to have more dark matter in
the simulations without baryons, a fact which we correct for using
the parameter $f$ which represents the fraction of total mass in the
form of baryons. The simulations use the value $f=0.15$ as suggested
by cosmology. The relationships between the total mass of the two
simulations are thus
\begin{eqnarray}
M^{\rm{DM}}(r_{200})&\simeq&M_{\rm{dm}}^S(r_{200})+M_b^S(r_{200})\\
(1-f)M^{\rm{DM}}(r_{200})&\simeq&M_{\rm{dm}}^S(r_{200}) .
\end{eqnarray}

The modified adiabatic contraction model is given by
\begin{equation}
M^{\rm{DM}}(\langle\bar{r}_i\rangle)r_i=\left[M_{\rm{dm}}^X(\langle\bar{r}_f\rangle)+M_b^S(\langle\bar{r}_f\rangle)\right]r_f\,,
\label{eq:gnedin}
\end{equation}
and the equation for the conservation of mass given by
\begin{equation}
(1-f) M^{\rm{DM}}(r_i)=M_{\rm{dm}}^X(r_f)\,. \label{eq:conserv}
\end{equation}
It is now possible for a given $A$ and $w$ (which will set the
$\langle\bar{r}_i\rangle$ dependency on $r_i$, as well as the
$\langle\bar{r}_f\rangle$ dependency on $r_f$) to use
Eqs.~(\ref{eq:gnedin}) and (\ref{eq:conserv}) to find
$M_{\rm{dm}}^X(r)$, using only $M^{\rm{DM}}(r)$ and $M_{b}^S(r)$, in
an attempt to reproduce $M_{\rm{dm}}^S(r)$. That is, we can solve
the above equations numerically to find $r_f$.  Using the above
model, we can hence determine predicted pinched dark matter halo
profiles for any given set of $A$ and $w$ values.

As discussed in the previous section, we use the rms dispersion of
the number of particles in radial bins across the five time frames
to estimate the uncertainties in our simulations. For each set of
($A$,$w$) values we perform the contraction of the dark-matter--only
halo, and compare that with the halo from the simulation containing
baryons. We restrict ourselves to the inner region of the dark
matter halo where the effects of baryonic contraction are most
prominent. Choosing the linear bin steps to be 1kpc (0.5kpc for the
smaller galaxies) smooths out most of the random substructure
density fluctuations and at the same time enables us to catch the
overall shape of the density profiles. To be more precise, we do the
$\chi^2$ analysis between the inner radius $r_{\rm{min}}$ and the
outer radius 0.1$r_{200}$ (i.e.\ 21 kpc for the two larger galaxies
and 10.5 kpc for the smaller), and divide the range into 19 linear
bins (consequently we have $N_{dof}=19-2=17$ degrees of freedom
(dof) in the fits).

\begin{figure}
\centerline{\epsfig{file=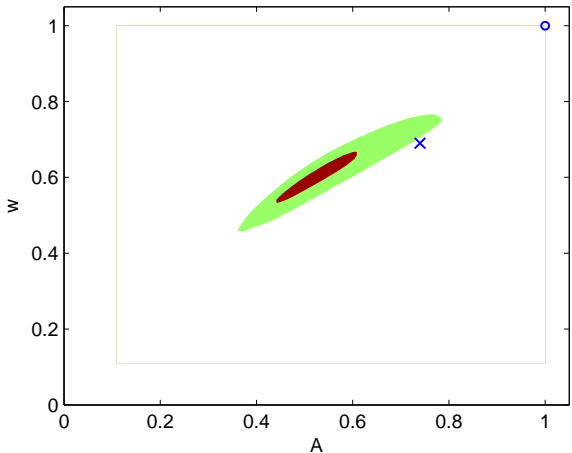,width=1\columnwidth}}
\caption{\it Best fit parameters for reconstructing the baryonic
compressed dark matter halo (S1) from its dark-matter--only halo
(DM1). The red/black area is the 1$\sigma$ (68\%) confidence region
and the green/gray is the 3$\sigma$ (99.7\%) confidence region. The
($A$,$w$) value, expected from the analysis of ellipticities as
proposed by Gnedin \emph{et al.} is marked by a cross, and the
original model, by Blumenthal \emph{et al.}, by a circle.}
\label{chisq15}
\end{figure}

\begin{figure}
\centerline{\epsfig{file=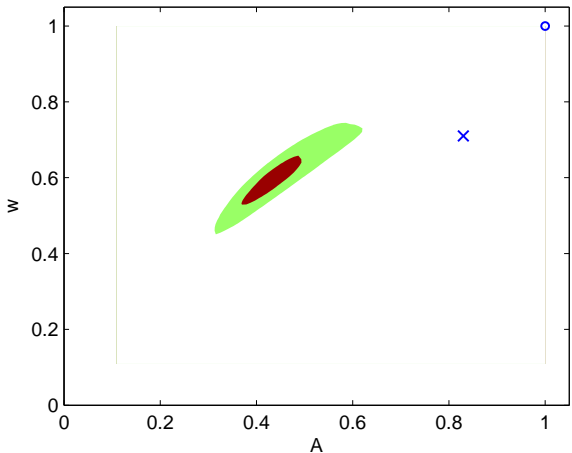,width=1\columnwidth}}
\caption{\it Best fit parameters for reconstructing the baryon
compressed dark matter halo (S2) from its dark-matter--only halo
(DM2). The red/black area is the 1$\sigma$ (68\%) confidence region
and the green/gray is the 3$\sigma$ (99.7\%) confidence region. The
($A$,$w$) value, expected from the analysis of ellipticities as
proposed by Gnedin \emph{et al.}, is marked by a cross, and the
original model, by Blumenthal \emph{et al.}, by a circle.}
\label{chisq18}
\end{figure}

To illustrate which ($A$,$w$) values provide good reconstructions of
the baryonically compressed dark-matter--only halo profiles, we show
in Fig.~\ref{chisq15} and \ref{chisq18} the $\chi^2$ of the
1$\sigma$ (68\% confidence) region and 3$\sigma$ (99.7\% confidence)
region in the $A$-$w$ plane. In other words inside the black/red
regions, $\chi^2$ is less than $\chi^2_{\rm{min}}$+2.3 and inside
the gray/green regions $\chi^2$ is less than
$\chi^2_{\rm{min}}$+11.8. For the canonical statistical scenario,
one expects a best $\chi^2$ per degree of freedom value of around 1,
which is also similar to what we find in our contour plots:
$\chi^2_{\rm{min}}/N_{dof} = 0.97, 0.77, 0.64(0.51)$ and $1.3$ for
halo 1, 2, 3(3-s) and 4, respectively. For any reasonable variation
of the number of bins and radii ranges in the inner region these
confidence levels stay rather stable and the best $\chi^2/N_{dof}$
stay fairly constant.

\begin{figure}
\centerline{\epsfig{file=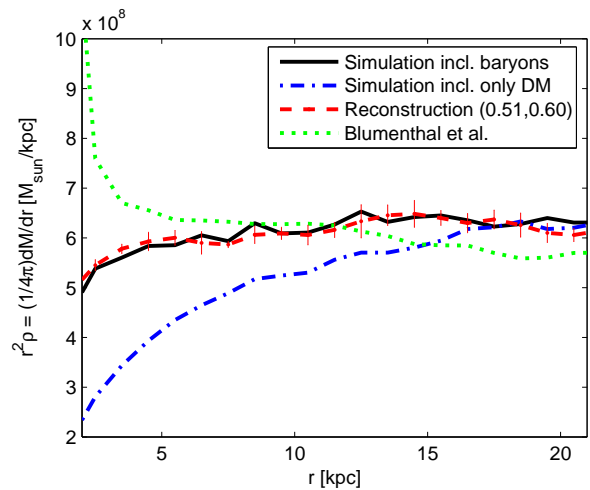,width=1\columnwidth}}
\caption{\it Density profiles for galaxy halo S1 and DM1. The
 blue/dot-dashed curve is $\rho r^2$ for the dark-matter--only
simulation, and the black/solid is $\rho r^2$ for the simulation
with baryons. The green/dotted curve shows the reconstruction using
the model of Blumenthal \emph{et al.} (i.e.\ ($A$, $w$)=(1,1)).
 The red/dotted curve with the error bars shows the best fit reconstruction, corresponding
to ($A$,$w$)=(0.51,0.60).} \label{fig:badgood}
\end{figure}

From the contour plots it follows that the fits for ($A$,$w$)=(1,1)
-- which corresponds to circular orbits and therefore the original
model of Blumenthal \emph{et al.} -- are significantly worse than
the fits for the optimal values. This can also be seen very clearly
by plotting the contracted $\textrm{d}M/\textrm{d}r$ profile using
the Blumenthal \emph{et al.} model and comparing this to the
contraction observed with the best fit values. For example, for
galaxy 1 the best fit value are ($A,w)\sim(0.5,0.6)$ and the
comparison are shown in Fig.~\ref{fig:badgood}.

The values of ($A$,$w$) obtained directly from the relation between
$\langle\bar{r}\rangle$ and $r$ in Eq.~(\ref{eq:poly}) are
significantly different from what is found by the above procedure.
This strongly suggests (not surprisingly), that there is more
physics at work than can be described by a simple two-parameter
model.

From our procedure to determine the best reconstruction of the
contracted dark matter halo, it should be obvious that the
confidence regions in Fig.~\ref{chisq15} and \ref{chisq18} should
not be interpreted as strict confidence regions for some correct
values of $A$ and $w$. Although the 68\% and 99.7\% confidence
levels in the figures correspond to the correct increase of
$\chi^2$, these confidence regions should rather be thought of
simply as a separation between values in the A-w plane which produce
good reconstruction of the baryonic contracted halo and those which
give a worse reconstruction. Moreover, there may not even be any
direct physical interpretation of the $A$ and $w$ found this way.

\begin{figure}
\centerline{\epsfig{file=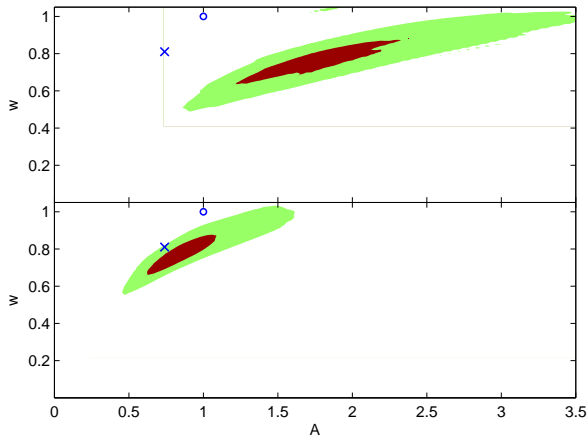,width=1\columnwidth}}
\caption{\it Best fit parameters for reconstructing the baryon
compressed dark matter halo from its dark-matter--only simulation,
using two different types of stellar feedback. Upper panel is using
simulation S3, whereas the lower is for an identical simulation, but
with stronger stellar feedback (S3-s). The red/black area is the
1$\sigma$ (68\%) confidence region and the green/gray is the
3$\sigma$ (99.7\%) confidence region. The stronger feedback
(producing a less concentrated galaxy) results in a significant
change of the best fit values towards smaller A.}
\label{fig:feedback}
\end{figure}

We also investigated the adiabatic contraction of the two smaller
galaxies 3 and 4.  Again, we found that the relationship between $r$
and $\langle\bar{r}\rangle$ in the dark matter halos of these
galaxies is modeled well by the power law in Eq.~($\ref{eq:poly}$).
For these simulations the best value of ($A$,$w$) for contraction
are substantially larger, and the preferred $A$ is actually even
larger than those predicted from the ellipticities. To elaborate
further on this point an extra simulation of the smaller galaxy S3
was performed where stronger early stellar energy feedback was
implemented at a level comparable with the top of the range
considered in \cite{astro-ph-0204366}). This extra simulation we
label \mbox{S3-s}.

It was found that increasing the feedback strength (and therefore
obtaining a less massive and somewhat more extended central galaxy)
did change the best fit values of ($A$,$w$) for baryonic contraction
considerably, as seen in Fig.~\ref{fig:feedback}. This suggests that
the details of the feedback, and its effect upon the concentration
of the baryons, is an important ingredient for predicting the
relationship between the final baryonic and dark matter density
profiles. Hence, we propose that it is not only the orbital
structure of the dark matter halo together with the final baryonic
profile which determines the final contracted profile of the dark
matter halo.

Summarizing, it is found that the parametric approach with different
($A$,$w$) in Eq.~(\ref{eq:poly}), (\ref{eq:gnedin}) and
(\ref{eq:conserv}) is able to very well reproduce the pinched dark
matter profiles. However, the ($A$,$w$) values are not universal,
and do not in general coincide with those predicted from the
ellipticity of the dark matter in Eq.~(\ref{eq:rbar}) and
(\ref{eq:poly}). Moreover, the details of the stellar feedback can
change which values of $A$ and $w$ are preferred in the modified
adiabatic compression model.

\section{Indirect dark matter detection}\label{sec:WIMP}

One immediate application of these results is the effect upon the
expected indirect signal from dark matter in the form of weakly
interacting massive particles (WIMP) annihilating in the galactic
center \cite{hep-ph-0002126}.
Several authors have tried to take into account the effect of
modified contraction models such as the one proposed by Gnedin
\emph{et al.}\ upon the expected number of annihilations from the
galactic center region (see e.g.\
\cite{astro-ph-0401512,astro-ph-0504631,hep-ph-0506204}).
It is today impossible for galaxy size simulations to get anywhere
near the length resolution corresponding to the very center of the
galaxy.  Nevertheless, we proceed in the spirit of comparison with
the existing literature by extrapolating our results into extremely
small radii.

We perform baryonic contraction of two different initial dark matter
profiles with a semirealistic spiral galaxy baryon profile, taking
some typical parameters from the Milky Way.  To model the Milky Way
baryon density we assume cylindrical symmetry, ignoring the
possibility of any bar. For the central bulge of stars we assume a
density of the form $\rho\propto r^{-\gamma}e^{-r/\lambda}$ while
for the disk we assume a Kuzmin profile. The Kuzmin disk can be
thought of as a delta function of matter in the $z$ direction ($z$
is the coordinate perpendicular to the disk) with a surface density
$\sigma_{\mathrm{disk}}(r)=\frac{cM_{\mathrm{disk}\infty}}{2\pi\left(r^2+c^2\right)^{\frac{3}{2}}}$
, where $M_{\mathrm{disk}\infty}$ is the total mass of the disk. We
choose the parameters of the model to match observations of the
Milky Way: $\gamma=1.85$, $\lambda=1 \mathrm{kpc}$, $c=5
\mathrm{kpc}$ and with the total disk and bulge mass
$M_{\mathrm{disk}\infty} = 5
M_{\mathrm{bulge}}=6.5\times10^{10}M_{\odot}$
\cite{astro-ph-9512064,astro-ph-9612059,astro-ph-0110390,Kent-1991me}.

The first dark matter profile we choose to contract is the standard
NFW profile, i.e. $(\alpha,\beta,\gamma)=(1,3,1)$ with a scale
radius of 20 kpc and a local density (i.e.\ at $r$=8.0 kpc) of
\mbox{0.3 GeV cm$^{-3}$}.

\begin{figure}
\begin{center}
\includegraphics[width=1\columnwidth]{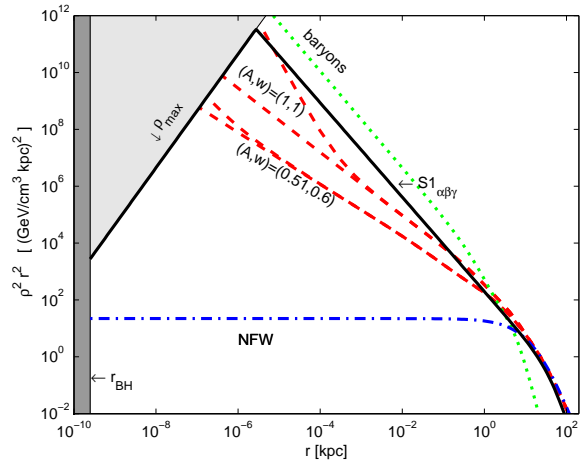}
\caption{\label{fig:NFWpinch}\it Diagram showing the contraction of
a standard NFW dark matter density profile (dot-dashed blue line) by
a baryon profile (dotted light green line) as described in the text.
The resulting dark matter profile (dashed red lines) are plotted for
$(A,w)=(1,1)$ and $(0.51,0.6)$, each splitting into two at low
radii, the denser corresponding to the density profile achieved from
a baryon profile which includes a central black hole.  The
extrapolated density profile from Table~\ref{tab:abg} for the dark
matter in simulation S1 is shown for comparison (solid black line).
Also shown are the maximum density line and the radius corresponding
to the lowest stable orbit around the central black hole.}
\end{center}
\end{figure}

We use the baryon profile described above to contract the dark
matter profile for some of the different values of ($A$,$w$) found
earlier in the paper and then calculate how the expected flux from
dark matter annihilations change.  The results of these contractions
can be seen in Fig.~\ref{fig:NFWpinch}, which show both the result
if a $2.6\times10^6$ $\mathrm{M}_\odot$ central supermassive black
hole is included in the baryonic profile and if it is not.

\begin{figure}
\begin{center}
\includegraphics[width=1\columnwidth]{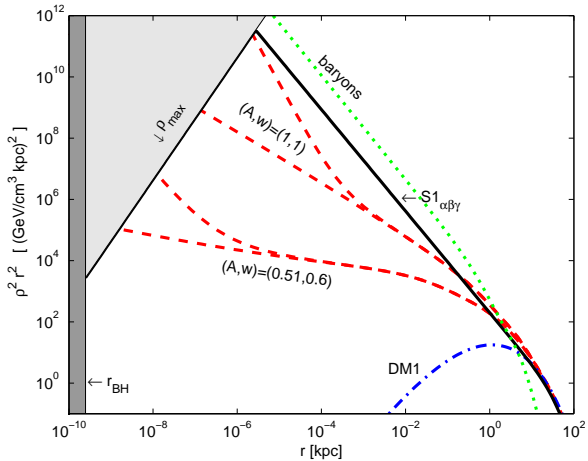}
\caption{\label{fig:DM1pinch}\it The same as
Fig.~\ref{fig:NFWpinch}, but where the initial dark matter profile
instead is the exponential profile with a continuously decreasing
logarithmic slope. The initial profile parameters are from the best
fit to the halo simulation DM1 as given in Table~\ref{tab:NFW_new}.}
\end{center}
\end{figure}

The same procedure is performed when the initial dark matter profile
instead is an exponential profile, where the logarithmic slope is
becoming continuously shallower, i.e.\ the profile in
Eq.~(\ref{eq:NFW_new}). We here use the best fit values found for
the DM1 halo as given in Table~\ref{tab:NFW_new} and again normalize
the local density to be $\rho_l\sim$ \mbox{0.3 GeV cm$^{-3}$}. The
results are presented in Fig.~\ref{fig:DM1pinch}.

We do not attempt to model the complicated dynamics at subparsec
scales of the galaxy
\cite{gondolosilk,astro-ph-0101481,astro-ph-0501555} other than
trying to take into account the maximum density due to
self-annihilations. In other words, a galactic dark matter halo
which has survived unperturbed by mergers or collisions for a time
scale $\tau_{gal}$ can not contain stable regions with densities
larger than $\rho_{max}\sim m_{\rm{dm}}/\langle\sigma v\rangle
\tau_{gal}$. We assume $\tau_{gal}=5\times 10^{9}$ years and for the
WIMP properties we adopt a dark matter mass of $m_{dm}=1$ TeV and an
annihilation cross section of $\langle\sigma v\rangle=3\times
10^{-26}$cm$^{3}$s$^{-1}$.

\begin{table}
\caption[]{\label{tab:annih}\it Estimated luminosity in erg s$^{-1}$
from dark matter annihilations in the center of contracted galaxies
for different contraction parameters ($A$,$w$) from an initial NFW
or the DM1 (in Table~\ref{tab:NFW_new}) dark matter profile and a
baryon profile including the super massive black hole, as described
in the text. We quota values for the inner 10 pc and 100 pc. Note
this is the total luminosity and not that in the form of some
specific particle species such as photons.}\vspace{0.0cm}
\begin{tabular*}{\columnwidth}{@{\extracolsep{\fill}} @{\hspace{3pt}}l@{\hspace{7pt}} l@{\hspace{8pt}} @{\hspace{5pt}}c@{\hspace{10pt}} c@{\hspace{2pt}} @{\hspace{5pt}}c@{\hspace{10pt}} c@{\hspace{2pt}}}\hline\hline
            &       &             \multicolumn{2}{c}{NFW}   & \multicolumn{2}{c}{DM1}\\
 $A$        & $w$                   & $\rm L_{10pc}        $ & $\rm L_{100pc}        $ & $\rm L_{10pc}        $ & $\rm L_{100pc}$ \vspace{0.05cm}\\\hline
 \multicolumn{2}{c@{\hspace{5pt}}}{Initial profile }& $\rm 3.9\times10^{33}$ & $\rm 3.9\times 10^{34}$ & $\rm 2.7\times10^{31}$ & $\rm 4.7\times 10^{33}$\\
 1          & 1                     & $\rm 2.8\times10^{40}$ & $\rm 2.8\times 10^{40}$ & $\rm 9.8\times10^{39}$ & $\rm 9.8\times 10^{39}$\\
 0.51       & 0.6                   & $\rm 2.8\times10^{37}$ & $\rm 3.5\times 10^{37}$ & $\rm 7.9\times10^{35}$ & $\rm 3.5\times 10^{36}$\\
 0.85       & 0.8                   & $\rm 5.9\times10^{38}$ & $\rm 6.0\times 10^{38}$ & $\rm 2.6\times10^{37} $ & $\rm 3.3\times 10^{37}$\\
 0.74       & 0.69                  & $\rm 6.9\times10^{37}$ & $\rm 7.8\times 10^{37}$ & $\rm 1.5\times10^{36}$ & $\rm 5.0\times 10^{36}$\\
\hline\hline
\end{tabular*}
\end{table}

In Table~\ref{tab:annih} we have looked at how the different values
of ($A$,$w$) found in the previous section would affect the dark
matter signal from the center of our idealized Milky Way. The
Blumenthal \emph{et al.} estimate gives fluxes far in excess of the
other estimates. The other three sets of ($A$,$w$) in
Table~\ref{tab:annih} are, respectively, those typical of the best
fit values found to reconstruct the larger galaxies, the best fit
values obtained when reconstructing the smaller \mbox{S3-s} galaxy,
and the best fit to the $r$ vs $\langle\bar{r}\rangle$ curves which
represent the ellipticities of the orbits in halo DM1.

All our values of $(A,w)$, when used to contract the dark matter
halos using our analytical baryonic density profile, give large
enhancements compared to the emission expected from the usual NFW
profile. This is true even when the initial dark matter distribution
follows the exponential profile as in Eq.~(\ref{eq:NFW_new}) which
does not initially posses a cusp. From the Table~\ref{tab:annih} we
deduce that the boost of the luminosity compared to the standard NFW
profile takes values in the range $10^2$ to $10^4$ for our
($A,w$)=(0.51,0.6) depending on the initial profile. The reader
should however note that these extrapolations to very small radii
neglect extra effects such as the scattering of dark matter
particles on stars or a noncentralized supermassive black hole
\cite{gondolosilk,astro-ph-0101481,astro-ph-0501555}.

The total luminosity does depend upon the ($A$,$w$) value, showing
that the flux which can be expected from dark matter annihilation
depends upon the extra physics included in simulations containing
baryonic hydrodynamics.

\section{Nonsphericity}\label{sec:non-sph}

In this section we relax the assumption of spherical symmetry and
determine the triaxial structure of the dark matter halos, assuming
ellipsoidal symmetry.

It is straightforward to obtain the moment of inertia tensor
\begin{equation}
 I_{ij} = \sum_k \left(r_k^2\delta_{ij}-r_{i,k} r_{j,k} \right)m_k
\end{equation}
of the dark matter, gas or stars, by summing over the masses ($m_k$)
and positions ($r_k$) of a matter component inside a spherical (or
ellipsoidal) shell of a given (major) radius $R$. By diagonalizing
$I_{ij}$ we find both the orientation of the semiaxes and the its
three eigenvalues $I_i$.  The three principal axes $(a,b,c)$ are
then found from the relation
\begin{equation}
a^2=  f_R\cdot\left(-I_{a}+I_{b}+I_{c}\right)
\end{equation}
and cyclic permutations thereof, where $f_R$ is a constant the
precise value of which depends upon the radial profile. To obtain an
accurate result we then repeat the search for the three axes
iteratively, in each step using only particles inside an elliptical
shell with semiaxes R, (b/a)R and (c/a)R as given by the previous
step. Further iterative steps are then taken until the results
converge. The principal axes are ordered such that $a\geq b\geq c$.

Having obtained the semiaxes, one determines whether a halo is
prolate, in other words shaped like a rugby ball (or an American
football), or oblate, i.e.\ flattened like a Frisbee, by introducing
the parameters $e=1-b/a$ and $f=1-c/a$.  If the halo is oblate or
flat it means that $a$ and $b$ are of similar size and much larger
than $c$ and the measure $T < 0.5$ where $T$ is defined as
\begin{equation}
T=\frac{a^2-b^2}{a^2-c^2}.
\end{equation}
Similarly, if $T > 0.5$ then the halo is prolate.

\begin{figure}
\centerline{\epsfig{file=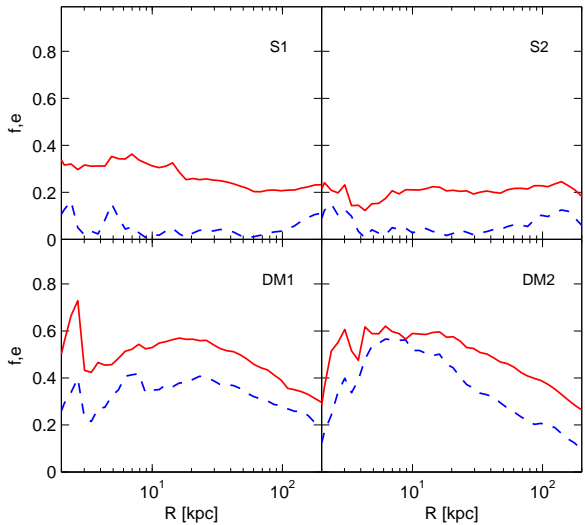,width=1\columnwidth}}
\caption{\it  The triaxial parameters $e=1-b/a$ (blue/dashed line)
and $f=1-c/a$ (red/solid line) of the dark matter halos for the two
larger galaxies with baryons (labeled S1 \& S2) and without baryons
(labeled DM1 \& DM2).  For a perfect oblate shape, $e=0$ and $f> 0$
whereas for a perfect prolate shape $e=f > 0$.} \label{fig:elllarge}
\end{figure}

\begin{figure}
\centerline{\epsfig{file=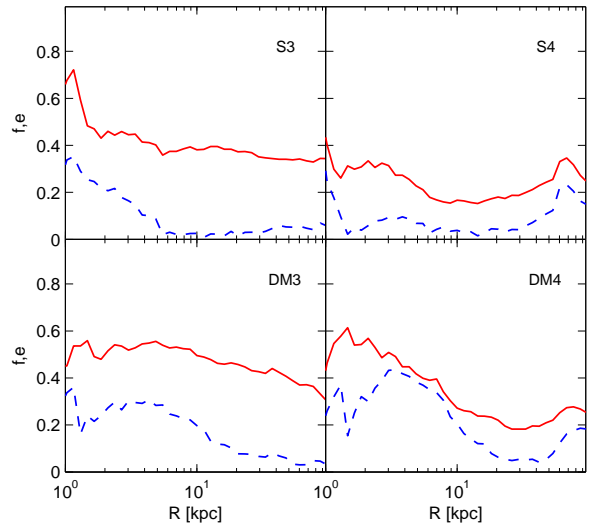,width=1\columnwidth}}
\caption{\it The triaxial parameters $e=1-b/a$ (blue dashed line)
and $f=1-c/a$ (red solid line) of the dark matter halos for the two
smaller galaxies with baryons (labeled S3\& S4) and without baryons
(labeled DM3 \& DM4).  For a perfect oblate shape, $e=0$ and $f> 0$
whereas for a perfect prolate shape $e=f > 0$.} \label{fig:ellsmall}
\end{figure}

The quantities $e$ and $f$ for the dark matter halos are shown in
Fig.~\ref{fig:elllarge} and \ref{fig:ellsmall} (as a function of the
major axis $R$). It is clear that the halos change from being
somewhat prolate ($e\sim f>0$) to being somewhat oblate ($e\sim0$,
$f>0$), when baryons are included in the simulations.  The values of
$T$ for the dark matter halos with and without baryons are listed in
Table~\ref{Tvalues}. In the presence of baryons the dark matter halo
tends to form an oblate halo, whereas when there are no baryons the
dark matter halo is more prolate.

\begin{table}
\caption[]{\it Values of the oblate/prolate-parameter $T$ inside 10
kpc for the halo simulations with and without
baryons.}\vspace{0.0cm}
\begin{tabular*}{\columnwidth}{@{\extracolsep{\fill}} l@{\hspace{20pt}} l@{\hspace{10pt}} l@{\hspace{10pt}} l@{\hspace{10pt}} l}\hline\hline
 Galaxy& 1 & 2& 3 & 4 \\
\hline
Sim. with baryons  & 0.076 & 0.15  & 0.048  & 0.24 \\
Sim. without baryons & 0.74  & 0.92 & 0.48  & 0.78 \\
\hline\hline
\end{tabular*}
\label{Tvalues}
\end{table}

\begin{figure*}
\centerline{\epsfig{file=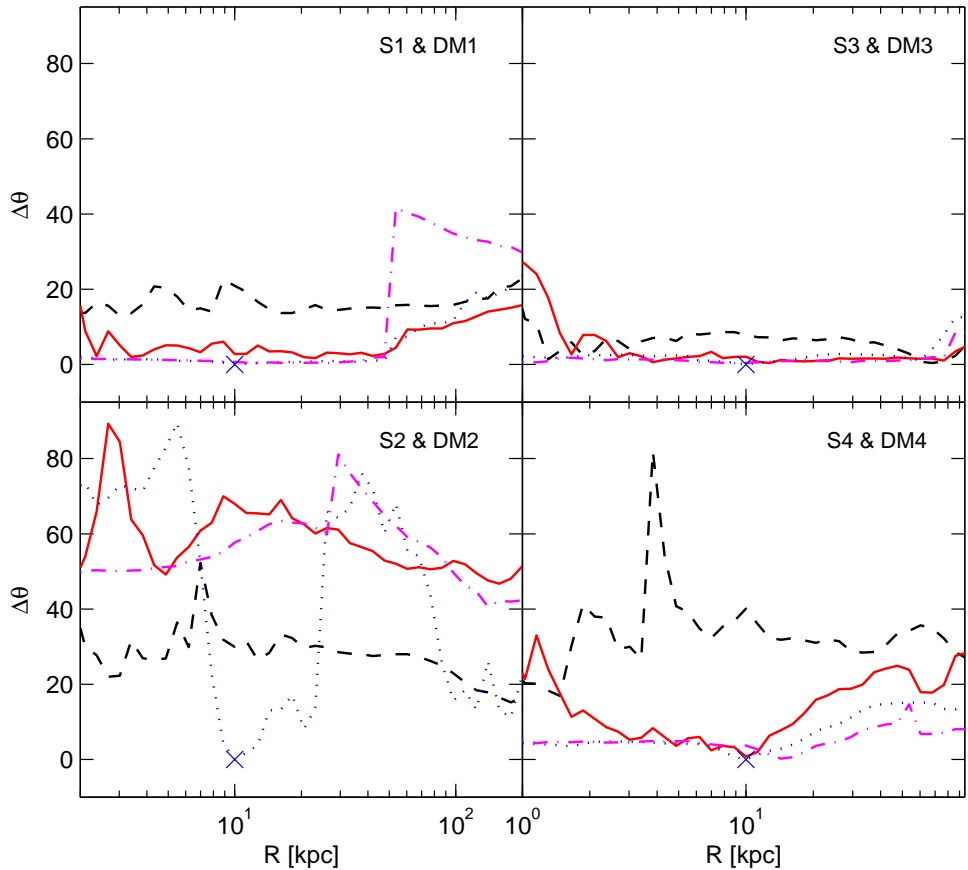,width=1.7\columnwidth}}
\caption{\it  Diagram showing angular alignment of the gas, stars
and dark matter in our four galaxy simulations. The vertical scale
is the difference in angle between the orientation of the smallest
axis (around which the moment of inertia is the greatest) of the
component in question relative to the axis of the gas inside 10 kpc
(by definition zero and marked with a blue cross). The blue dotted
line is the gas, the magenta dot-dashed line are the stars and the
red solid line is the dark matter in the simulation with baryons.
The black dashed line is the dark matter in the simulation without
baryons, showing that the baryonic disk is aligned with the plane of
the original density distribution.} \label{fig:angle}
\end{figure*}

Given these results, the obvious thing to check is whether the
principal axes of the dark matter distributions and the baryon
distributions are aligned. Figure \ref{fig:angle} shows the
alignment between the stellar disk, the gaseous disk and the dark
matter ``disk''.  The diagram is obtained by finding the orientation
vectors of the minor axis ($c$) for different lengths scales (i.e.
for different values of the major axis $R$ of the ellipsoidal used
in the calculation of the moment of inertia tensor $I_{ij}$). The
parameter $\Delta\theta$ is then the angle between each of these
vectors and a reference vector defined to correspond to the
orientation vector of the gaseous disk determined inside $R$ = 10
kpc. From the diagrams it follows that for galaxies 1, 3 and 4, the
orientation of the minor axis of the gas, stars and dark matter are
strongly correlated with each other. However, for galaxy 2 the there
is no clear alignment between the different matter components at all
(not even within the gas itself at different radii). The reason for
this discrepancy is likely due to that the baryonic galaxy 2 has
experienced a late-time merger, making it irregular.

Figures \ref{fig:elllarge}, \ref{fig:ellsmall} and \ref{fig:angle}
provide interesting information on the interaction between baryons
and dark matter in spiral galaxies. Figures \ref{fig:elllarge} and
\ref{fig:ellsmall} indicate that the formation of a disk by gas
cooling and contraction causes the dark matter halo to lose most of
its prolateness and instead become oblate, flattened slightly in the
disk plane. Inside a few kpc, the dark matter can tend to be less
oblate and instead develop a prolate structure. This is the case
for, e.g., the dark matter halo in simulation S3, in which it is
aligned with a strong stellar bar (see \cite{astro-ph-0506627} for a
dedicated study of central bar structures in the cold dark matter).

Additionally, Fig.~\ref{fig:angle} shows us that the orientation of
the baryonic disk is rather correlated with the orientation of the
flattest part of the dark matter halo in the simulation without
baryons.  The dark matter therefore seems to have a role in
determining the orientation of the baryonic disk. Subsequently, the
formation and contraction of the baryonic disk causes most of the
halo triaxiality to be erased, resulting in a somewhat flattened,
approximately cylindrically symmetric halo
\cite{astro-ph-9309001,astro-ph-0603487}.

The amount of triaxiality of dark matter halos is a generic
prediction in the hierarchial, cold dark matter model of structure
formation, and observational probes of halo shapes are therefore a
fundamental test of this model. Unfortunately, observational
determination of halo shapes is a difficult task, and only coarse
constraints exist. Probes of the Milky Way halo indicate that it
should be rather spherical with $f \lesssim 0.2$ and that an oblate
structure of $f \sim 0.2$ might be preferable (see, e.g.,
\cite{astro-ph-0608343} and references therein).
 Milky Way sized halos formed in dissipationless simulations are
usually predicted to be considerably more triaxial and prolate,
although a large scatter is expected
\cite{Frenk88,Dubinski:1991bm,Warren:1992tr,Thomas98,astro-ph-0202064,Bailin:2004wu,Bett:2006zy}.
Including dissipational baryons into the numerical simulation, and
thereby converting the halo prolateness into a slightly oblate halo,
might turn out to be essential to produce good agreement with
observations. In a similar numerical study \cite{astro-ph-0405189}
the effect of the baryons on the halo shape was also found to
washout the triaxiality. However, in their one realization of a
Milky Way size halo no clear oblateness of the galactic halo was
recognized, but rather an almost spherical halo was achieved (with
$f \sim e \sim 0.1$).

Having determined the ellipsoidal triaxiality of the dark matter
distribution, one can include this information in the profile fits.
To exemplify this, we refit the halo of S1 (the galaxy found to be
most similar to the Milky Way) to the '$\alpha\beta\gamma$' profile
as given in Eq.~(\ref{eq:abg}), but now with the replacement
\begin{equation}\label{abg_tri}
    r \rightarrow \tilde{r} = \sqrt{x^2+\frac{y^2}{(1-e)^2} + \frac{z^2}{(1-f)^2}}.
\end{equation}
The best fit values are given in Table~\ref{tab:abg_tri}, where we
have used axis ratios as found within $R$=10 kpc (from
Fig.~\ref{fig:elllarge} we note that we are quite insensitive to the
exact choice of $R$).

\begin{table}
\begin{center}
 \caption[]{\it Best fit parameters for the ellipsoidal triaxial
dark matter halo from simulation S1 (including baryons), using model
Eq.~(\ref{eq:abg}).}\vspace{0.0cm}
\begin{tabular*}{\columnwidth}{@{\extracolsep{\fill}} lcccccc}\hline\hline
 f      & e  & $r_{s}$ [kpc]& $\alpha$ & $\beta$ & $\gamma$ & $\chi^2_{dof}$ [46 dof]\\\hline
 0.31   &0.02&  44.6   & 2.02     & 3.17    & 1.86     &  1.2
 \\\hline\hline
\end{tabular*}
\label{tab:abg_tri}
\end{center}
\end{table}

The ($\alpha,\beta,\gamma$) parameters of the ellipsoidal triaxial
fit do not change much compared to the spherically symmetric fit,
this is because the flattening of the dark matter halo is not very
strong. However, the important difference is that the amount of
flattening of the dark matter halo in the galactic plane is taken
into account, and in fact, the goodness of the fit is slightly
improved.

The oblate structure of the dark matter will also have some effects
on the expected indirect dark matter signal \cite{astro-ph-0504631}.
However, the baryonic pinching effect does not produce such highly
flattened halo profiles as the one proposed in, e.g.,
\cite{astro-ph-0508617} to explain the EGRET observed diffuse gamma
excess by WIMP annihilations (see also the critique in
\cite{astro-ph-0602632} on the dark matter interpretation of the
EGRET signal).

\section{Summary and Conclusions}\label{sec:summary}

We have presented results comparing the structure of galactic dark
matter halos formed in N-body simulations including only dark matter
to that of {\it the same} halos formed in N-body/hydrodynamical
simulations of galaxies, containing dark matter, stars and gas. From
our selected high resolution galaxies three out of the four galaxies
formed in the hydrodynamical simulations contain very distinct disks
of gas and stars, and central stellar bulges, the fourth is strongly
barred.

The central slope of the dark matter density profiles becomes
significantly steeper when baryons are present, with the average
logarithmic slope parameter $\gamma=-\textrm{d} \log \rho
/\textrm{d}\log r$ increasing from 1.3$\pm$0.2 to 1.9$\pm$0.2 at
about 1\% of the virial radius.

The pinching of dark matter halos in response to the cooling and
contraction of baryons was investigated further, to test adiabatic
contraction models for the case where galaxies of realistic linear
sizes and other properties are formed.

In relation to the orbital structure of dark-matter--only halos, the
mean of the time averaged radius $\langle\bar{r}\rangle$ of dark
matter particles versus radius $r$ is very well described by a power
law relation (specified by two parameters $A$ and $w$) as suggested
by Gnedin \emph{et al}.
 Moreover, it is found that the Gnedin \emph{et al.}\
\cite{astro-ph-0406247} prescription for adiabatic contraction is
much more successful at reproducing the density profile of dark
matter in the simulations with baryons, than the standard scenario
of Blumenthal \emph{et al.} \cite{ASJOA.301.27}, in which circular
orbits are assumed.  However, the parameters of the Gnedin \emph{et
al.}\ model, ($A$,$w$), which give the best fit for the baryonic
contraction, are somewhat different from the ($A$,$w$) parameters
describing the averaged orbital structure of the dark-matter--only
halos. Given the ($A$,$w$) uncertainty estimates, described in the
text, this difference appears to be significant. In addition, it is
also found that the contraction reconstruction values of ($A$,$w$)
also depend on the strength of the stellar feedback in simulations
of otherwise identical halos, further indicating (perhaps as one
might expect) that the effects of baryonic pinching are more
complicated, than what can be captured in this two-parameter model.

Our results indicate that the amount of baryonic pinching of the
dark matter halos are overestimated in earlier works applying the
adiabatic compression model by Blumenthal \emph{et al.}, at least
for Milky Way sized disk galaxies. This has ramifications for
predictions of the (putative) dark matter annihilation flux from the
galactic center. It is found, that the flux can be reduced by
several orders of magnitude, although baryonic contraction still
boosts the signal significantly above the value one would expect on
the basis of simulations containing only dark matter.

Finally, we have determined the triaxiality of the dark matter
halos. Dark matter only halos were found to be significantly more
prolate than halos containing baryons.  The influence from baryons
flatten the dark matter into a slightly oblate halo
($c/a=0.73\pm0.11$) aligned in the same plane as the stellar/gaseous
disk.  Moreover, in the simulations containing baryons, galactic
disks tend to form in the planes aligned with the flattest parts of
the dark matter halos formed in the corresponding simulation without
baryons.

{\bf Acknowledgments} We are extremely grateful for conversations
with Lars Bergstr\"om, Anne Green, Anatoly Klypin, Edvard
M\"ortsell, George Rhee and Christian Walck. All computations
reported in this paper were performed on the IBM SP4 and SGI Itanium
II facilities provided by Danish Center for Scientific Computing
(DCSC). MF receives funding from the Swedish Research Council
(Vetenskapsr\aa det). The Dark Cosmology Centre is funded by the
DNRF.

\bigskip


\end{document}